\documentclass[aps,prd,twocolumn,amsmath,amssymb,showpacs,superscriptaddress]{revtex4}

\usepackage[draft]{hyperref}
\usepackage{amssymb}
\usepackage{amsmath}
\usepackage{graphicx}
\usepackage{bm}

\input{tcilatex}

\begin{document}

\title{Surface polaritons of a left-handed curved slab}
\author{Mounir Faddaoui}
\email{faddaoui@univ-corse.fr}
\affiliation{UMR CNRS 6134 SPE, Equipe Physique Th\'eorique, \\
Universit\'e de Corse, Facult\'e des Sciences, Bo{\^\i}te Postale
52, 20250 Corte, France}
\author{Antoine Folacci}
\email{folacci@univ-corse.fr}
\affiliation{UMR CNRS 6134 SPE, Equipe Physique Th\'eorique, \\
Universit\'e de Corse, Facult\'e des Sciences, Bo{\^\i}te Postale
52, 20250 Corte, France}
\author{Paul Gabrielli}
\email{gabrieli@univ-corse.fr}
\affiliation{UMR CNRS 6134 SPE, Equipe Ondes et Acoustique, \\
Universit\'e de Corse, Facult\'e des Sciences, Bo{\^\i}te Postale 52, 20250
Corte, France}

\begin{abstract}
We study the propagation of surface polaritons in a left-handed curved slab,
i.e. a curved slab made of a negative-refractive-index material. We consider
the effects of the slab curvature on their dispersion relations and
attenuations. We show more particularly that surface polaritons with a
\textquotedblleft left-handed behavior\textquotedblright\ (i.e. with
opposite group and phase velocities) propagate without any attenuation.
\end{abstract}

\maketitle



\section{Introduction}

Recently, motivated by theoretical considerations developed a long time ago
by Veselago \cite{Veselago} and following insights from Pendry and coworkers
\cite{Pendry96,Pendry98,Pendry99}, Shelby, Schultz, Smith and colleagues
\cite{Smith2000a,Smith2001a,Smith2001b} have been able to build, for the
first time, by combining arrays of wires and split-ring resonators, an
artificial medium for which the electric permittivity, the magnetic
permeability and therefore the refractive index are simultaneously negative
in the microwave frequency range. They have thus opened a new era for optics
because in negative-refractive-index media (also called left-handed
materials or double-negative media), electromagnetism presents unusual
properties due to various anomalous effects such as reversed Doppler shift,
reversed Cerenkov radiation, negative radiation pressure and inverse
Snell-Descartes law \cite{Veselago,PendrySmith2004}. Since then, several
other groups have successfully fabricated left-handed media and it is now
possible to test and to exploit \textquotedblleft left-handed
electromagnetism" in a large range of frequencies. Of course, the unusual
and remarkable properties of negative-refractive-index media could
revolutionize optics, optoelectronics and communications and, as a
consequence, this recent and rapidly evolving new field of physics has
attracted interest of many researchers and many technological applications
are already considered including superlenses, band-pass filters, beam
guiders, light-emitting devices, cloaking devices ...

In this article, we shall focus our attention on a particular problem of
left-handed electromagnetism namely the propagation of surface polaritons in
a curved slab made of a negative-refractive-index material and we shall
study the effects of curvature on their dispersion relations and
attenuations. Surface polaritons (and other guided modes) supported by a
left-handed flat slab have been studied extensively in the recent years
(see, e.g., \cite%
{RuppinPLA2001,Shadrivov2003,RaoOng2003,ParkLeeFuZhang2005,Shadrivov2005,
HeCaoShen2005,Shadrivov2006,DarmanyanKobyakovChowdhury2007,MoreauFelbacq2008,WangLi2008}%
) due to the central role that they seem to play in the superlensing
phenomenon \cite{Pendry00,Feise02,Haldane02,RaoOng2003} and in the giant
Goos-H{\"{a}}nchen effect \cite{ShadrivovEtAl03} as well as due to their
other potential applications, for example, if we have in mind the
development of unconventional photonic integrated circuits and ultra-compact
plasmon-based integrated circuits.

In the left-handed flat slab configuration, we can consider that the
properties of surface polaritons are now completely known. They have been
obtained from rather elementary calculations involving homogeneous and
inhomogeneous plane waves. To our knowledge, there exists no description of
surface polaritons guided by a left-handed curved slab despite the interest
of this problem. Indeed, these surface polaritons are necessary in order to
explain the resonant behavior of hollow left-handed nanoparticles \cite%
{Ruppin2005}\ or of coaxial cylindrical cables made of negative-index
metamaterials \cite{Kushwaha2010} and they could be used to transmit
efficiently information by using curved waveguides. Mathematically, the
description of surface polaritons guided by curved interfaces can be
achieved in the framework of complex angular momentum techniques or, in
other words, by using the Regge pole machinery \cite%
{Watson18,Sommerfeld49,Nus92,Grandy}. In optics, such techniques permit one
to naturally shed light on the physics lying behind the transcendental
equations involving non-elementary special functions which usually appear in
the description of surface waves propagating close to a curved interface.
Recently, these techniques have been used in order to describe surface
polaritons guided by single interfaces separating a dispersive medium and an
ordinary one (see \cite%
{BerryMV1975,AnceyDecaniniFolacciGabrielli1,AnceyDecaniniFolacciGabrielli2,
AnceyDecaniniFolacciGabrielli3,AnceyDecaniniFolacciGabrielli4,hasegawaNockelDeutsch1,
hasegawaNockelDeutsch2}%
). In the present article, we shall apply these techniques in order to
analyze the properties of surface polaritons guided in a left-handed
cylindrical slab.

Our paper is organized as follows. In Section 2, we briefly recall the
theory of surface polaritons guided in a left-handed flat slab embedded in
an ordinary dielectric medium. We note that these surface polaritons
propagate without any losses and we more particularly emphasize the
existence of frequency ranges where they present a \textquotedblleft
left-handed behavior", i.e. with opposite group and phase velocities. In
Section 3, we extend our analysis to the case of a left-handed cylindrical
slab and we consider the effects of its curvature on surface polariton
properties. We observe that curvature slightly modifies the dispersion
relations of the surface polaritons while, in general, it can lead to
important attenuations. However, it is worth pointing out that curvature
does not induce any losses for some surface polaritons and, in particular,
for those presenting a left-handed behavior. In Section 4, we recall the
main results of our work and briefly discuss some possible practical
applications.

It should be noted that, in our article, we implicitly assume the time
dependence $\exp \left( -i\omega t\right) $ for all the fields.

\section{Surface polaritons of a left-handed ordinary slab}

\subsection{General remarks and notations}

\begin{figure}[tbp]
\includegraphics[height=4.5cm,width=8.6cm]{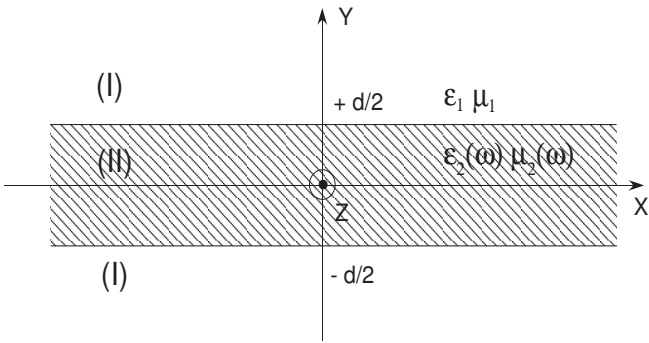}
\caption{Geometry of the left-handed slab.} \label{fig:Slab}
\end{figure}

We consider a symmetric slab of thickness $d$ made of a
negative-refractive-index material (region II) imbedded in a host medium
(region I) with electric permittivity $\varepsilon _{1}>0$ and magnetic
permeability $\mu _{1}>0$ both frequency independent (see Fig.~\ref{fig:Slab}
for the geometry of the system and the notations). As far as the electric
permittivity $\epsilon _{2}(\omega )$ and the magnetic permeability $\mu
_{2}(\omega )$ of the slab are concerned, we assume they are respectively
given by
\begin{equation}
\varepsilon _{2}(\omega )=1-\frac{\omega _{p}^{2}}{\omega ^{2}}  \label{eps2}
\end{equation}%
and
\begin{equation}
\mu _{2}(\omega )=1-\frac{F\omega ^{2}}{\omega ^{2}-\omega _{0}^{2}}%
=(1-F)\left( \frac{\omega ^{2}-\omega _{b}^{2}}{\omega ^{2}-\omega _{0}^{2}}%
\right)  \label{mu2}
\end{equation}%
where $0<F<1$ and $\omega _{b}=\omega _{0}/\sqrt{1-F}$. Of course, the
parameters $\omega _{p}$, $\omega _{0}$ and $F$ depend on the structure of
the negative-refractive-index material but we do not attribute any
\textquotedblleft microscopic" interpretation to them. We only assume that $%
\omega _{0}<\omega _{b}<\omega _{p}$. We then have $\epsilon (\omega )<0$ in
the frequency range $\omega \in \left] 0,\omega _{p}\right[ $ and $\mu
(\omega )<0$ in the frequency range $\omega \in \left] \omega _{0},\omega
_{b}\right[ $. Thus, the electric permittivity, the magnetic permeability
and the refractive index are simultaneously negative in the region $\omega
_{0}<\omega <\omega _{b}$. In that region the metamaterial presents a
left-handed behavior. Furthermore, in order to describe below wave
propagation, we also introduce the refractive indices
\begin{subequations}
\label{ind_n}
\begin{eqnarray}
&&n_{1}=\sqrt{\varepsilon _{1}\mu _{1}},  \label{ind_n1} \\
&&n_{2}\left( \omega \right) =\sqrt{\varepsilon _{2}\left( \omega \right)
\mu _{2}\left( \omega \right) },  \label{ind_n2}
\end{eqnarray}%
as well as the wave numbers
\end{subequations}
\begin{subequations}
\label{kappa}
\begin{eqnarray}
&&\kappa _{1}\left( \omega \right) =n_{1}\left( \frac{\omega }{c}\right) ,
\label{kappa1} \\
&&\kappa _{2}\left( \omega \right) =n_{2}\left( \omega \right) \left( \frac{%
\omega }{c}\right) .  \label{kappa2}
\end{eqnarray}

We shall study the guided modes propagating along the slab. Here and from
now on, we choose to treat our problem in a two-dimensional setting,
ignoring the $z$ coordinate. We shall consider separately the $H$ and $E$
polarizations. For the $H$ polarization, the magnetic field $\mathbf{H}$ is
parallel to the $z$ axis and, from Maxwell's equations, it is easy to show
that it satisfies the Helmholtz equation
\end{subequations}
\begin{subequations}
\label{HEquHz}
\begin{eqnarray}
&&\left[ \Delta +n_{2}^{2}(\omega )\left( \frac{\omega }{c}\right) ^{2}%
\right] H_{z}^{\mathrm{II}}(\mathbf{x})=0 \  \mathrm{for} \
-d/2<y<d/2, \nonumber \\
& & \label{HEqu1Hz} \\
&&\left[ \Delta +n_{1}^{2}\left( \frac{\omega }{c}\right) ^{2}\right] H_{z}^{%
\mathrm{I}}(\mathbf{x})=0 \  \mathrm{for} \ y<-d/2\text{ and
}y>d/2, \nonumber \\
& &\label{HEqu2Hz}
\end{eqnarray}%
where $\mathbf{x}=(x,y)$. From the continuity of the tangential components
of the electric and magnetic fields at the interface between regions I and
II, it can be shown that the $z$-component of the magnetic field satisfies,
for$\ y=\pm d/2$,
\end{subequations}
\begin{subequations}
\label{BCHz}
\begin{eqnarray}
&&H_{z}^{\mathrm{I}}(\mathbf{x})=H_{z}^{\mathrm{II}}(\mathbf{x}),
\label{BCHz1} \\
&&\frac{1}{\epsilon _{1}}\frac{\partial H_{z}^{\mathrm{I}}}{\partial n}(%
\mathbf{x})=\frac{1}{\epsilon _{2}(\omega )}\frac{\partial H_{z}^{\mathrm{II}%
}}{\partial n}(\mathbf{x}).  \label{BCHz2}
\end{eqnarray}%
For the $E$ polarization, the electric field $\mathbf{E}$ is parallel to the
$z$ axis and, from Maxwell's equations, we obtain the Helmholtz equation
\end{subequations}
\begin{subequations}
\label{HEquEz}
\begin{eqnarray}
&&\left[ \Delta +n_{2}^{2}(\omega )\left( \frac{\omega }{c}\right) ^{2}%
\right] E_{z}^{\mathrm{II}}(\mathbf{x})=0 \  \mathrm{for}\
-d/2<y<d/2, \nonumber \\
& &
\label{HEqu1Ez} \\
&&\left[ \Delta +n_{1}^{2}\left( \frac{\omega }{c}\right) ^{2}\right] E_{z}^{%
\mathrm{I}}(\mathbf{x})=0 \  \mathrm{for}\ y<-d/2\text{ and
}y>d/2. \nonumber \\
& & \label{HEqu2Ez}
\end{eqnarray}%
Due to the continuity of the tangential components of the electric and
magnetic fields at the interface between regions I and II, the $z$-component
of the electric field satisfies, for$\ y=\pm d/2$,
\end{subequations}
\begin{subequations}
\label{BCEz}
\begin{eqnarray}
&&E_{z}^{\mathrm{I}}(\mathbf{x})=E_{z}^{\mathrm{II}}(\mathbf{x}),
\label{BCEz1} \\
&&\frac{1}{\mu _{1}}\frac{\partial E_{z}^{\mathrm{I}}}{\partial n}(\mathbf{x}%
)=\frac{1}{\mu _{2}(\omega )}\frac{\partial E_{z}^{\mathrm{II}}}{\partial n}(%
\mathbf{x}).  \label{BCEz2}
\end{eqnarray}%
Finally, it should be noted that the mathematical solutions of the two
previous problems can be separated into even (or symmetric) and odd (or
antisymmetric) modes due to the symmetry of the slab under the
transformation $y\rightarrow -y.$

\subsection{Surface polaritons for the H polarization}

For the $H$ polarization the guided modes propagating in the slab are
solutions of the Helmholtz equation (\ref{HEquHz}) of the form
\end{subequations}
\begin{equation}
\mathbf{H}=\left\{
\begin{array}{ll}
H_{1}e^{ikx-\alpha y}\mathbf{e}_{z} & \qquad \mathrm{for}\quad y>\frac{d}{2},
\\
H_{2}e^{ikx}f_{\pm }\left( \beta y\right) \mathbf{e}_{z} & \qquad \mathrm{for%
}-\frac{d}{2}<y<\frac{d}{2}, \\
\pm H_{1}e^{ikx+\alpha y}\mathbf{e}_{z} & \qquad \mathrm{for}\quad y<-\frac{d%
}{2},%
\end{array}%
\right.  \label{SPplaque_pol_H}
\end{equation}%
with $f_{+}\left( \beta y\right) =\cosh \left( \beta y\right) $ and $%
f_{-}\left( \beta y\right) =\sinh \left( \beta y\right) $. In Eq. (\ref%
{SPplaque_pol_H}) the $+$ and $-$ signs are respectively associated with
even and odd modes. Here $k$ describes the propagation of the magnetic field
along the $x$ axis, while $\alpha $ and $\beta $ are functions permitting us
to describe its decay near the interfaces. By inserting Eq. (\ref%
{SPplaque_pol_H}) into Eqs. (\ref{HEquHz}) and (\ref{BCHz}), we obtain from
one hand
\begin{eqnarray}
{\alpha }^{2}\left( \omega ,k\right) &=&k^{2}-\kappa _{1}^{2}\left( \omega
\right) ,  \label{expalpha3} \\
{\beta }^{2}\left( \omega ,k\right) &=&\kappa _{2}^{2}\left( \omega \right)
-k^{2},  \label{expbeta3}
\end{eqnarray}%
and from the other hand
\begin{equation}
\frac{\alpha \left( \omega ,k\right) }{\beta \left( \omega ,k\right) }=-%
\frac{\varepsilon _{1}}{\varepsilon _{2}(\omega )}\tanh \left[ \frac{\beta
\left( \omega ,k\right) d}{2}\right] ,  \label{rel dis3}
\end{equation}%
\begin{equation}
H_{2}=H_{1}\frac{\exp \left[ -\frac{\alpha \left( \omega ,k\right) d}{2}%
\right] }{\cosh \left[ \frac{\beta \left( \omega ,k\right) d}{2}\right] },
\label{relH3}
\end{equation}%
for the even solutions, and
\begin{equation}
\frac{\alpha \left( \omega ,k\right) }{\beta \left( \omega ,k\right) }=-%
\frac{\varepsilon _{1}}{\varepsilon _{2}(\omega )}\coth \left[ \frac{\beta
\left( \omega ,k\right) d}{2}\right] ,  \label{rel dis4}
\end{equation}%
\begin{equation}
H_{2}=H_{1}\frac{\exp \left[ -\frac{\alpha \left( \omega ,k\right) d}{2}%
\right] }{\sinh \left[ \frac{\beta \left( \omega ,k\right) d}{2}\right] },
\label{relH4}
\end{equation}%
for the odd solutions. Equations (\ref{rel dis3}) and (\ref{rel dis4})
provide implicitly, for the $H$ polarization, the dispersion relations $%
k=k(\omega )$ or $\omega =\omega (k)$ for the guided modes propagating in
the slab. From now on, we shall only consider the guided surface modes. We
then require the following conditions
\begin{equation*}
{\alpha }^{2}\left( \omega ,k\right) >0\text{ and }{\beta }^{2}\left( \omega
,k\right) <0,
\end{equation*}%
which imply
\begin{subequations}
\label{c.exi3}
\begin{eqnarray}
k^{2} &>&\kappa _{1}^{2}\left( \omega \right) ,  \label{c.exi3a} \\
k^{2} &>&\kappa _{2}^{2}\left( \omega \right) ,  \label{c.exi3b}
\end{eqnarray}%
for the existence conditions of the surface polaritons.

\subsection{Surface polaritons for the E polarization}

For the $E$ polarization the guided modes propagating in the slab are
solutions of the Helmholtz equation (\ref{HEquEz}) of the form
\end{subequations}
\begin{equation}
\mathbf{E}=\left\{
\begin{array}{ll}
E_{1}e^{ikx-\alpha y}\mathbf{e}_{z} & \qquad \mathrm{for}\quad y>\frac{d}{2},
\\
E_{2}e^{ikx}f_{\pm }\left( \beta y\right) \mathbf{e}_{z} & \qquad \mathrm{for%
}-\frac{d}{2}<y<\frac{d}{2}, \\
\pm E_{1}e^{ikx+\alpha y}\mathbf{e}_{z} & \qquad \mathrm{for}\quad y<-\frac{d%
}{2},%
\end{array}%
\right.  \label{SPplaque_pol_E}
\end{equation}%
with $f_{+}\left( \beta y\right) =\cosh \left( \beta y\right) $ and $%
f_{-}\left( \beta y\right) =\sinh \left( \beta y\right) $. Here $k,$ $\alpha
$ and $\beta $ as well as the $+$ and $-$ signs keep their previous
interpretations. Substituting Eq. (\ref{SPplaque_pol_E}) into Eq. (\ref%
{HEquEz}) provides again the relations (\ref{expalpha3}) and (\ref{expbeta3}%
). Moreover, by inserting (\ref{SPplaque_pol_E}) into (\ref{BCEz}), we
obtain
\begin{equation}
\frac{\alpha \left( \omega ,k\right) }{\beta \left( \omega ,k\right) }=-%
\frac{\mu _{1}}{\mu _{2}(\omega )}\tanh \left[ \frac{\beta \left( \omega
,k\right) d}{2}\right] ,  \label{rel dis7}
\end{equation}%
\begin{equation}
E_{2}=E_{1}\frac{\exp \left[ -\frac{\alpha \left( \omega ,k\right) d}{2}%
\right] }{\cosh \left[ \frac{\beta \left( \omega ,k\right) d}{2}\right] },
\label{relH7}
\end{equation}%
for the even solutions, and
\begin{equation}
\frac{\alpha \left( \omega ,k\right) }{\beta \left( \omega ,k\right) }=-%
\frac{\mu _{1}}{\mu _{2}(\omega )}\coth \left[ \frac{\beta \left( \omega
,k\right) d}{2}\right] ,  \label{rel dis8}
\end{equation}%
\begin{equation}
E_{2}=E_{1}\frac{\exp \left[ -\frac{\alpha \left( \omega ,k\right) d}{2}%
\right] }{\sinh \left[ \frac{\beta \left( \omega ,k\right) d}{2}\right] },
\label{relH8}
\end{equation}%
for the odd solutions. Equations (\ref{rel dis7}) and (\ref{rel dis8})
provide implicitly, for the $E$ polarization, the dispersion relations $%
k=k(\omega )$ or $\omega =\omega (k)$ for the guided modes propagating in
the slab and we still have the existence conditions (\ref{c.exi3}) for the
surface polaritons.

\begin{figure}[tbp]
\includegraphics[height=6.5cm,width=8.6cm]{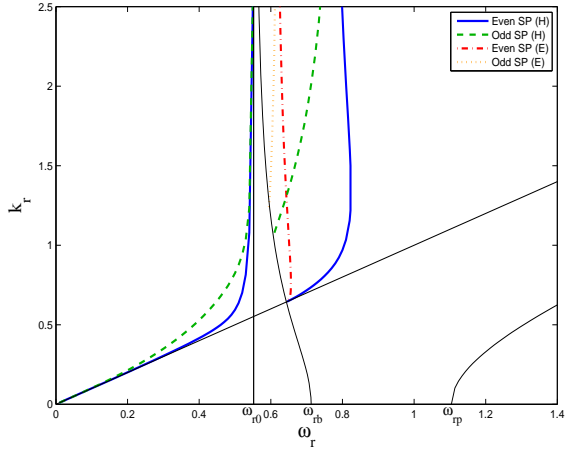}
\caption{Dispersion relations of the surface polaritons \ for a
left-handed
flat slab embedded in vacuum ($\protect\varepsilon _{1}=1,$ $\protect\mu %
_{1}=1$). The tiny curves delimit the regions in which surface
polaritons can exist (cf. Eqs. (\protect\ref{c.exi3})).}
\label{fig:dispers_plaque}
\end{figure}
\begin{figure}[tbp]
\includegraphics[height=6.5cm,width=8.6cm]{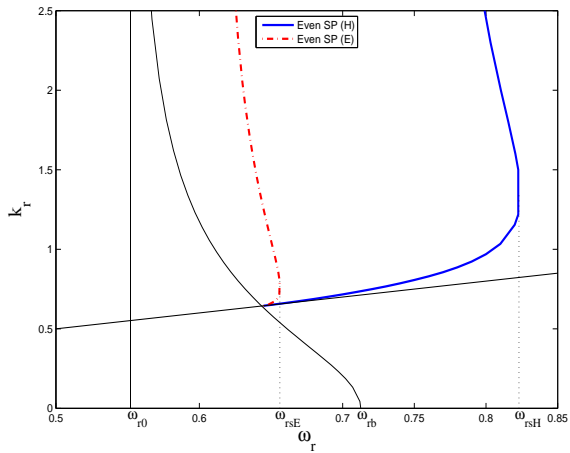}
\caption{Zoom in on dispersion relations of the even surface
polaritons with
``left-handed behavior''. The left-handed flat slab is embedded in vacuum ($%
\protect\varepsilon _{1}=1,$ $\protect\mu _{1}=1$).}
\label{fig:dispers_plaque_zoom}
\end{figure}

\subsection{Numerical aspects}

In Fig. \ref{fig:dispers_plaque} we display the surface polariton dispersion
relations for the slab embedded in vacuum ($\varepsilon _{1}=1$ and $\mu
_{1}=1$). They are plotted in the form $k_{r}=k_{r}\left( \omega _{r}\right)
$ where $k_{r}$ and $\omega _{r}$ are the reduced parameters defined by
\begin{subequations}
\label{par_red}
\begin{eqnarray}
k_{r} &=&\frac{kd}{c},  \label{par_red_1} \\
\omega _{r} &=&\frac{\omega d}{c}.  \label{par_red_2}
\end{eqnarray}
\end{subequations}
As far as the characteristics of the left-handed material are
concerned, we work with $F=0.4$ and with the reduced frequencies
$\omega _{r0}=\omega _{0}d/c=0.552$, $\omega _{rb}=\omega
_{b}d/c\approx 0.7127$ and $\omega _{rp}=\omega _{p}d/c=1.104$. Even
though we restrict ourselves to that particular configuration, the
results we obtain numerically are in fact very general and they
permit us to correctly illustrate the theory. Similarly, the global
aspects of the dispersion curves are rather independent of the value
of $\varepsilon _{1}.$

Our results are in agreement with those already obtained in the literature
by several authors (see e.g. Ref. \cite{RuppinPLA2001}). So, we shall not
lengthily analyze them. However we would like to emphasize the slope
inversion (as well as its consequences) which occurs for the two branches
corresponding to the even surface polaritons in the frequency range $\omega
_{r0}<\omega _{r}<\omega _{rp}$. Indeed, let us denote by $\omega _{rsH}$
and $\omega _{rsE}$ the slope inversion frequencies of the even surface
polaritons for the $H$ and $E$ polarizations (see Fig. \ref%
{fig:dispers_plaque_zoom}). In the $H$ polarization case, for $\omega _{r}$
below but near $\omega _{rsH}$ there exist two values for the reduced
propagation constant $k_{r}$ corresponding to two distinct behavior for the
surface polaritons:\newline
(i) for the lower $k_{r}$ value, the slope is positive and, as a
consequence, the group and phase velocities are both positive and the
surface polariton presents an ordinary behavior,\newline
(ii) for the higher $k_{r}$ value, the slope is negative and, as a
consequence, the group and phase velocities are opposite and the surface
polariton presents a left-handed behavior.\newline
Such a result will have\ a crucial importance for surface polaritons
propagating in a curved slab. In the $E$ polarization case, similar
considerations apply with $\omega _{rsE}$ replacing $\omega _{rsH}.$

\section{Surface polaritons of a left-handed curved slab}

\begin{figure}[tbp]
\includegraphics[height=7.60cm,width=7.51cm]{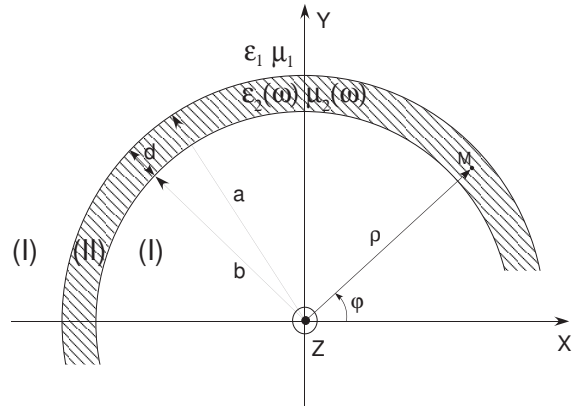}
\caption{Geometry of the left-handed curved slab.} \label{fig:Curved
Slab}
\end{figure}

\subsection{General remarks and notations}

We assume that the previous slab of thickness $d=a-b$ has been bent and
occupies the region corresponding to the range $b<\rho <a$ in the usual
cylindrical coordinate system $\left( \rho ,\varphi ,z\right) $ (see Fig. %
\ref{fig:Curved Slab}).

We shall study the guided modes propagating along the slab by
treating again our problem in a two-dimensional setting (ignoring
the $z$ coordinate) and considering separately the $H$ and $E$
polarizations. For the $H$ polarization, the magnetic field
$\mathbf{H}$ is parallel to the $z$ axis and, from Maxwell's
equations, it is easy to show that it satisfies the Helmholtz
equation
\begin{subequations}
\label{HEquHz_curved}
\begin{eqnarray}
&&\left[ \Delta +n_{2}^{2}(\omega )\left( \frac{\omega }{c}\right) ^{2}%
\right] H_{z}^{\mathrm{II}}(\mathbf{x})=0\quad \mathrm{for}\ b<\rho <a,
\label{HEquHz1_curved} \\
&&\left[ \Delta +n_{1}^{2}\left( \frac{\omega }{c}\right) ^{2}\right] H_{z}^{%
\mathrm{I}}(\mathbf{x})=0\quad \mathrm{for}\text{ }0<\rho <b\text{ and }\rho
>a,  \nonumber \\
&& \label{HEquHz2_curved}
\end{eqnarray}%
\end{subequations}
where $\mathbf{x}=(\rho ,\varphi )$. From the continuity of the
tangential components of the electric and magnetic fields at the
interface between regions I and II, we can show that the relations
(\ref{BCHz}) remain valid
for $\rho =a$ and $\rho =b.$ For the $E$ polarization, the electric field $%
\mathbf{E}$ is parallel to the $z$ axis and, from Maxwell's
equations, we obtain the Helmholtz equation
\begin{subequations}
\label{HEquEz_curved}
\begin{eqnarray}
&&\left[ \Delta +n_{2}^{2}(\omega )\left( \frac{\omega }{c}\right) ^{2}%
\right] E_{z}^{\mathrm{II}}(\mathbf{x})=0\quad \mathrm{for}\ b<\rho <a,
\label{HEquEz1_curved} \\
&&\left[ \Delta +n_{1}^{2}\left( \frac{\omega }{c}\right) ^{2}\right] E_{z}^{%
\mathrm{I}}(\mathbf{x})=0\quad \mathrm{for}\text{ }0<\rho <b\text{ and }\rho
>a.  \nonumber \\
&& \label{HEquEz2_curved}
\end{eqnarray}%
\end{subequations}
Due to the continuity of the tangential components of the electric
and magnetic fields at the interface between regions I and II, the
$z$-component of the electric field again satisfies the relations
(\ref{BCEz}) for$\ \rho =a$ and $\rho =b$. Finally, it should be
noted that the mathematical solutions of the two previous problems
cannot be naturally separated into even (or symmetric) and odd (or
antisymmetric) modes due to the symmetry breaking induced by the
curvature of the slab.

\subsection{Surface polaritons for the H polarization}

For the $H$ polarization the guided modes propagating in the slab are
solutions of the Helmholtz equation (\ref{HEquHz_curved}) which can be
expressed in terms of Bessel functions \cite{AS65} on the form $\mathbf{H}%
=H_{z}\mathbf{e}_{z}$ with
\begin{equation}
H_{z}(\rho ,\varphi )=\left\{
\begin{array}{ll}
& A_{1}H_{\lambda }^{(1)}\left[ \kappa _{1}\left( \omega \right)
\rho \right]
e^{i\lambda \varphi }   \quad \mathrm{for}\,\rho >a, \\
& \left( A_{2}J_{\lambda }[\kappa _{2}\left( \omega \right) \rho
]+B_{2}H_{\lambda }^{(1)}[\kappa _{2}\left( \omega \right) \rho
]\right)
e^{i\lambda \varphi }   \\
& \qquad\qquad\qquad\qquad\qquad  \mathrm{for}\,b<\rho <a, \\
& A_{3}J_{\lambda }[\kappa _{1}\left( \omega \right) \rho
]e^{i\lambda \varphi
}   \qquad \mathrm{for}\,0<\rho <b.%
\end{array}%
\right. \label{SPcurved_slab_pol_H}
\end{equation}%
Here $\lambda $ is an azimuthal complex constant describing the
propagation along the curved slab while $\kappa _{1}\left( \omega
\right) $ and $\kappa _{2}\left( \omega \right) $ are still defined
by Eqs. (\ref{kappa}). Substituting (\ref{SPcurved_slab_pol_H}) into
(\ref{BCHz}) provides a system of four equations with four unknowns
that can be expressed in the matrix form
\begin{equation}
M^{H}\left( \lambda ,\omega \right) \text{\thinspace }C^{H}=0
\label{syst_mat_H}
\end{equation}%
with
\begin{widetext}
\begin{eqnarray}
&&M^{H}\left( \lambda ,\omega \right) =  \left(
\begin{array}{cccc}
H_{\lambda }^{(1)}[\kappa _{1}\left( \omega \right) a] & -J_{\lambda
}[\kappa _{2}\left( \omega \right) a] & -H_{\lambda }^{(1)}[\kappa
_{2}\left( \omega \right) a] & 0 \\
\sqrt{\frac{\mu _{1}}{\varepsilon _{1}}}H_{\lambda }^{(1)^{\prime }}[\kappa
_{1}\left( \omega \right) a] & -\sqrt{\frac{\mu _{2}(\omega )}{\varepsilon
_{2}(\omega )}}J_{\lambda }^{\prime }[\kappa _{2}\left( \omega \right) a] & -%
\sqrt{\frac{\mu _{2}(\omega )}{\varepsilon _{2}(\omega )}}H_{\lambda
}^{(1)^{\prime }}[\kappa _{2}\left( \omega \right) a] & 0 \\
0 & -\sqrt{\frac{\mu _{2}(\omega )}{\varepsilon _{2}(\omega )}}J_{\lambda
}^{\prime }[\kappa _{2}\left( \omega \right) b] & -\sqrt{\frac{\mu
_{2}(\omega )}{\varepsilon _{2}(\omega )}}H_{\lambda }^{(1)^{\prime
}}[\kappa _{2}\left( \omega \right) b] & \sqrt{\frac{\mu _{1}}{\varepsilon
_{1}}}J_{\lambda }^{\prime }[\kappa _{1}\left( \omega \right) b] \\
0 & -J_{\lambda }[\kappa _{2}\left( \omega \right) b] & -H_{\lambda
}^{(1)}[\kappa _{2}\left( \omega \right) b] & J_{\lambda }[\kappa _{1}\left(
\omega \right) b]%
\end{array}%
\right)     \label{matrice_H}
\end{eqnarray}%
\end{widetext}
and
\begin{equation}
C^{H}=\left(
\begin{array}{c}
A_{1} \\
A_{2} \\
B_{2} \\
A_{3}%
\end{array}%
\right) .  \label{mat_CH}
\end{equation}%

The system (\ref{syst_mat_H}) admits non trivial solutions $C^{H}$
only if
\begin{equation}
\det M^{H}\left( \lambda ,\omega \right) =0.  \label{rel_disp_curved_H}
\end{equation}
The complex solutions $\lambda $ of Eq. (\ref{rel_disp_curved_H})
are the so-called Regge poles and they can be interpreted as complex
angular momenta \cite{Watson18,Sommerfeld49,Nus92,Grandy}. The
relations $\lambda =\lambda \left( \omega \right) $ obtained from
(\ref{rel_disp_curved_H}) provide the dispersion relation as well as
the attenuation of all the guided modes propagating in the slab.
Among all these modes, there exist guided modes corresponding to
those already present in the ordinary slab, as well as an infinity
of new modes due to the slab curvature (whispering gallery modes).
Of course, we shall focus our attention only to the guided modes
associated with the surface polaritons described in Section 2.

\subsection{Surface polaritons for the E polarization}

For the $E$ polarization the guided modes propagating in the slab are
solutions of the Helmholtz equation (\ref{HEquEz_curved}) of the form $%
\mathbf{E}=E_{z}\mathbf{e}_{z}$ with
\begin{equation}
E_{z}(\rho ,\varphi )=\left\{
\begin{array}{ll}
& A_{1}H_{\lambda }^{(1)}[\kappa _{1}\left( \omega \right) \rho
]e^{i\lambda
\varphi }   \quad   \mathrm{for}\,\rho >a, \\
& \left( A_{2}J_{\lambda }[\kappa _{2}\left( \omega \right) \rho
]+B_{2}H_{\lambda }^{(1)}[\kappa _{2}\left( \omega \right) \rho
]\right)
e^{i\lambda \varphi }  \\
& \qquad\qquad\qquad\qquad\qquad    \mathrm{for}\,b<\rho <a, \\
& A_{3}J_{\lambda }[\kappa _{1}\left( \omega \right) \rho
]e^{i\lambda \varphi
}   \qquad  \mathrm{for}\,0<\rho <b.%
\end{array}%
\right.  \label{SPcurved_slab_pol_E}
\end{equation}%
Substituting (\ref{SPcurved_slab_pol_E}) into (\ref{BCEz}) provides a system
of four equations with four unknowns that can be expressed in the matrix form
\begin{equation}
M^{E}\left( \lambda ,\omega \right) \text{\thinspace }C^{E}=0
\label{syst_mat_E}
\end{equation}%
with%
\begin{widetext}
\begin{eqnarray}
&&M^{E}\left( \lambda ,\omega \right) =  \left(
\begin{array}{cccc}
H_{\lambda }^{(1)}[\kappa _{1}\left( \omega \right) a] & -J_{\lambda
}[\kappa _{2}\left( \omega \right) a] & -H_{\lambda }^{(1)}[\kappa
_{2}\left( \omega \right) a] & 0 \\
\sqrt{\frac{\varepsilon _{1}}{\mu _{1}}}H_{\lambda }^{(1)^{\prime }}[\kappa
_{1}\left( \omega \right) a] & -\sqrt{\frac{\varepsilon _{2}(\omega )}{\mu
_{2}(\omega )}}J_{\lambda }^{\prime }[\kappa _{2}\left( \omega \right) a] & -%
\sqrt{\frac{\varepsilon _{2}(\omega )}{\mu _{2}(\omega )}}H_{\lambda
}^{(1)^{\prime }}[\kappa _{2}\left( \omega \right) a] & 0 \\
0 & -\sqrt{\frac{\varepsilon _{2}(\omega )}{\mu _{2}(\omega )}}J_{\lambda
}^{\prime }[\kappa _{2}\left( \omega \right) b] & -\sqrt{\frac{\varepsilon
_{2}(\omega )}{\mu _{2}(\omega )}}H_{\lambda }^{(1)^{\prime }}[\kappa
_{2}\left( \omega \right) b] & \sqrt{\frac{\varepsilon _{1}}{\mu _{1}}}%
J_{\lambda }^{\prime }[\kappa _{1}\left( \omega \right) b] \\
0 & -J_{\lambda }[\kappa _{2}\left( \omega \right) b] & -H_{\lambda
}^{(1)}[\kappa _{2}\left( \omega \right) b] & J_{\lambda }[\kappa _{1}\left(
\omega \right) b]%
\end{array}%
\right)   \label{matrice_E}
\end{eqnarray}%
\end{widetext}
and%
\begin{equation}
C^{E}=\left(
\begin{array}{c}
A_{1} \\
A_{2} \\
B_{2} \\
A_{3}%
\end{array}%
\right) .  \label{mat_CE}
\end{equation}

\begin{figure}[tbp]
\includegraphics[height=6.5cm,width=8.6cm]{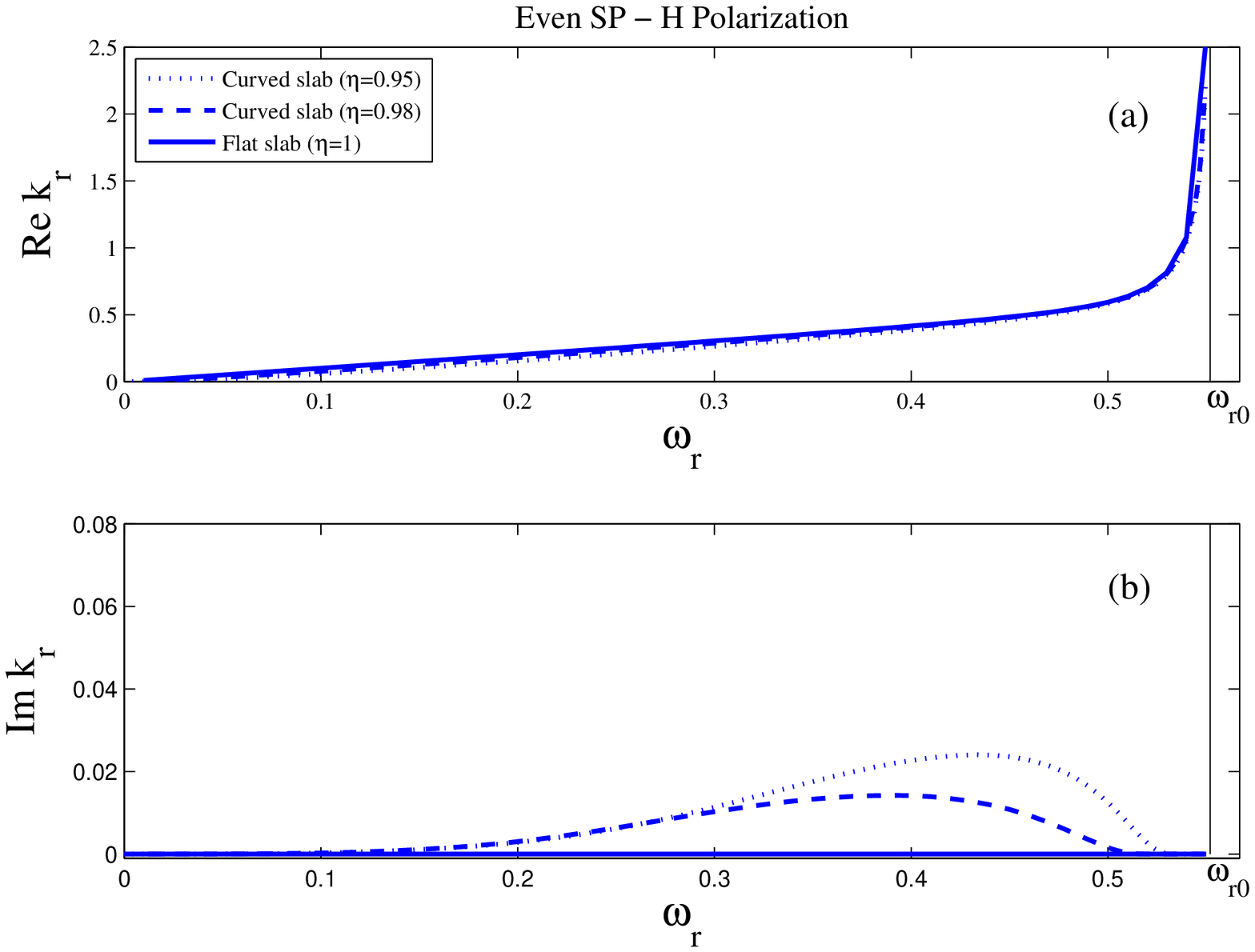}
\caption{Dispersion relations (a) and attenuations (b) of surface
polaritons
guided in left-handed flat and curved slabs embedded in vacuum ($\protect%
\varepsilon _{1}=1,$ $\protect\mu _{1}=1$): role of the slab
curvature. \
Even surface polariton, $H$ polarization, frequency range $0<\protect\omega %
_{r}<\protect\omega _{r0}$.} \label{fig:EvenSP_polH_BF}
\end{figure}
\begin{figure}[tbp]
\includegraphics[height=6.5cm,width=8.6cm]{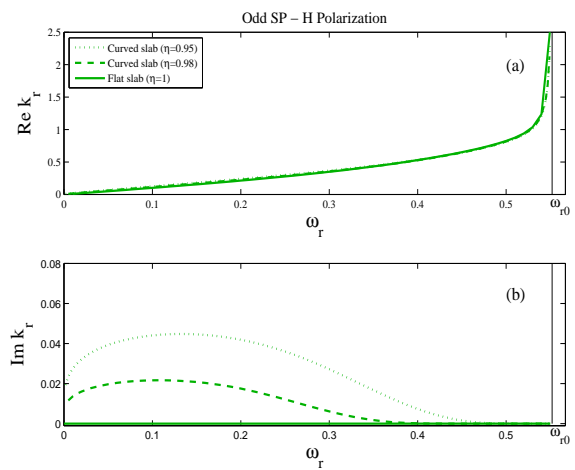}
\caption{Dispersion relations (a) and attenuations (b) of surface
polaritons
guided in left-handed flat and curved slabs embedded in vacuum ($\protect%
\varepsilon _{1}=1,$ $\protect\mu _{1}=1$): role of the slab
curvature. \
Odd surface polariton, $H$ polarization, frequency range $0<\protect\omega %
_{r}<\protect\omega _{r0}$.} \label{fig:OddSP_polH_BF}
\end{figure}

The system (\ref{syst_mat_E}) admits non trivial solutions $C^{E}$ only if%
\begin{equation}
\det M^{E}\left( \lambda ,\omega \right) =0  \label{rel_disp_curved_E}
\end{equation}%
which provides the dispersion relation and the attenuation of all the guided
modes propagating in the slab. Similarly, we shall focus our attention only
to the guided modes associated with the surface polaritons described in
Section 2.

\subsection{Numerical aspects}

We shall restrict our numerical study to cylindrical slabs of weak
curvature. In other words, we shall assume that $d \ll a$ and $d \ll
b$ and therefore that $\eta = b/a \approx 1$. This hypothesis
permits us to consider that surface polaritons propagate very close
to $\rho =a$ and thus the arc length they cover is given by ${\cal
L}=a \varphi$. As a consequence, the complex wave number $k$
describing propagation along the curved slab is linked to the
complex angular momentum $\lambda$ by $k=\lambda/a$ [in Eqs.
(\ref{SPcurved_slab_pol_H}) and (\ref{SPcurved_slab_pol_E}), we can
write $\exp (i \lambda \varphi)=\exp (i k {\cal L})$] and therefore
$\mathrm{Re}\ k=\mathrm{Re}\ \lambda/a$ and $\mathrm{Im}\
k=\mathrm{Im}\ \lambda/a$ provide respectively the dispersion
relation and the attenuation of surface polaritons.

\begin{figure}[tbp]
\includegraphics[height=6.5cm,width=8.6cm]{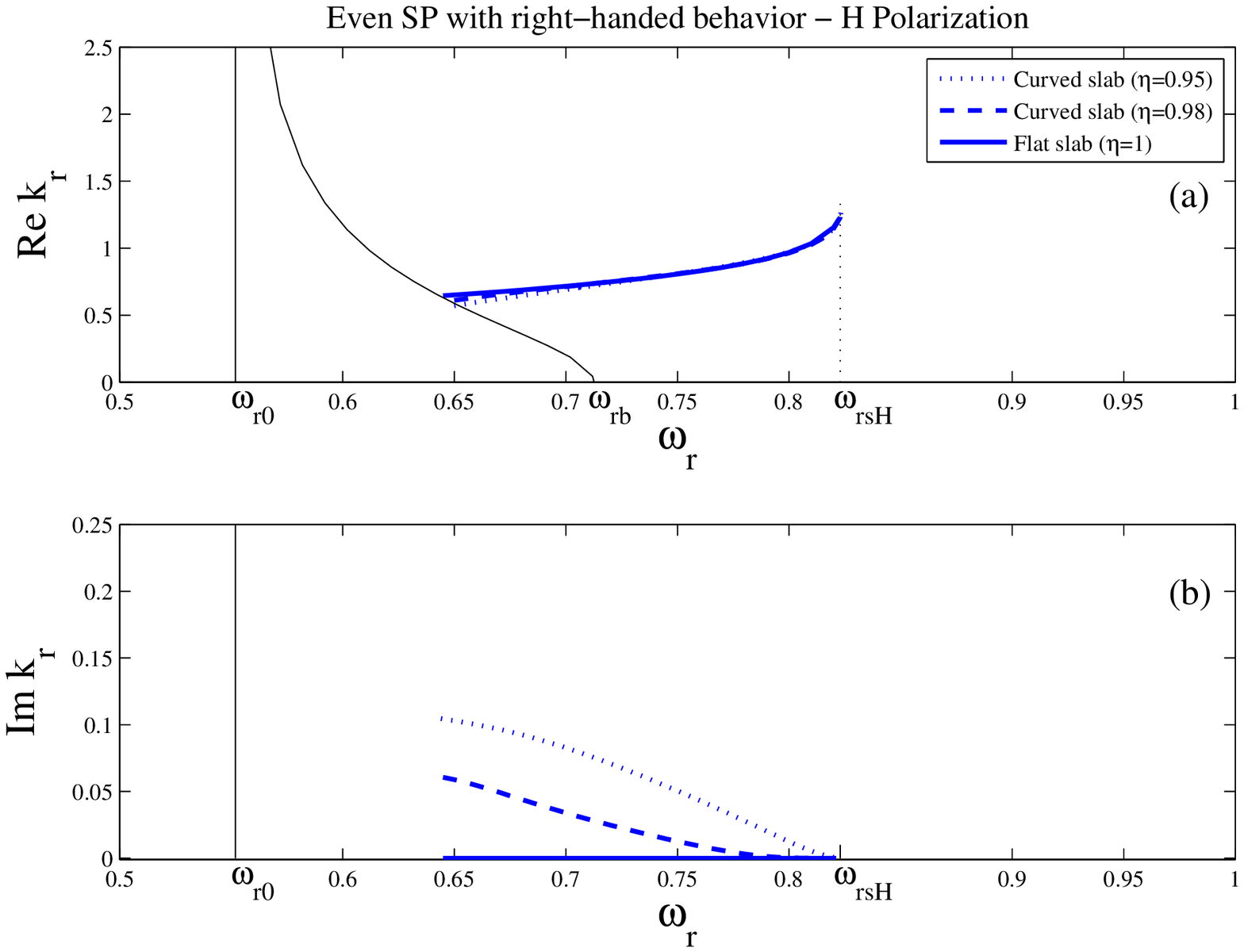}
\caption{Dispersion relations (a) and attenuations (b) of surface
polaritons
guided in left-handed flat and curved slabs embedded in vacuum ($\protect%
\varepsilon _{1}=1,$ $\protect\mu _{1}=1$): role of the slab
curvature. \ Even surface polariton with right-handed behavior, $H$
polarization, frequency range $\protect\omega _{r}>\protect\omega
_{r0}$.} \label{fig:EvenSP_rh_polH_HF}
\end{figure}
\begin{figure}[tbp]
\includegraphics[height=6.5cm,width=8.6cm]{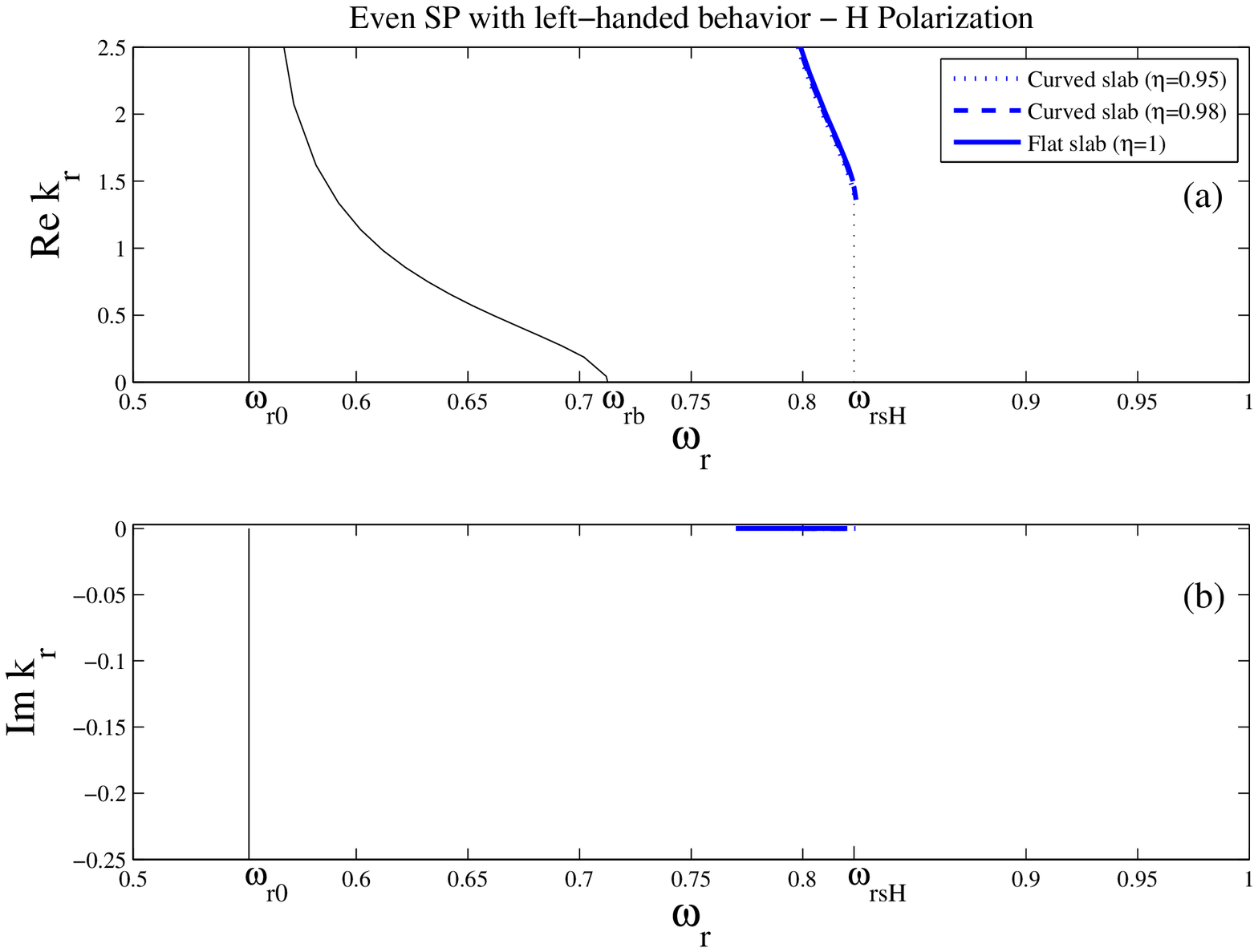}
\caption{Dispersion relations (a) and attenuations (b) of surface
polaritons
guided in left-handed flat and curved slabs embedded in vacuum ($\protect%
\varepsilon _{1}=1,$ $\protect\mu _{1}=1$): role of the slab
curvature. \ Even surface polariton with left-handed behavior, $H$
polarization, frequency range $\protect\omega _{r}>\protect\omega
_{r0}$.} \label{fig:EvenSP_lh_polH_HF}
\end{figure}

\begin{figure}[tbp]
\includegraphics[height=6.5cm,width=8.6cm]{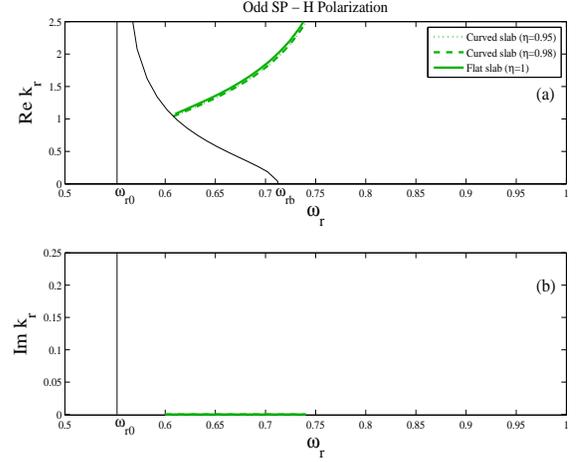}
\caption{Dispersion relations (a) and attenuations (b) of surface
polaritons
guided in left-handed flat and curved slabs embedded in vacuum ($\protect%
\varepsilon _{1}=1,$ $\protect\mu _{1}=1$): role of the slab
curvature. \
Odd surface polariton, $H$ polarization, frequency range $\protect\omega %
_{r}>\protect\omega _{r0}$.} \label{fig:OddSP_polH_HF}
\end{figure}

\begin{figure}[tbp]
\includegraphics[height=6.5cm,width=8.6cm]{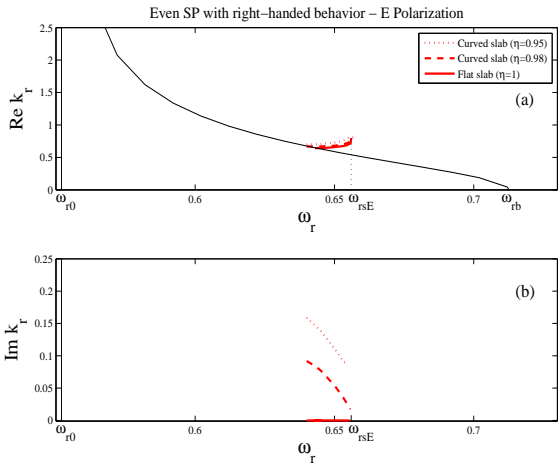}
\caption{Dispersion relations (a) and attenuations (b) of surface
polaritons
guided in left-handed flat and curved slabs embedded in vacuum ($\protect%
\varepsilon _{1}=1,$ $\protect\mu _{1}=1$): role of the slab
curvature. \ Even surface polariton with right-handed behavior, $E$
polarization, frequency range $\protect\omega _{r}>\protect\omega
_{r0}$.} \label{fig:EvenSP_rh_polE_HF}
\end{figure}

\begin{figure}[tbp]
\includegraphics[height=6.5cm,width=8.6cm]{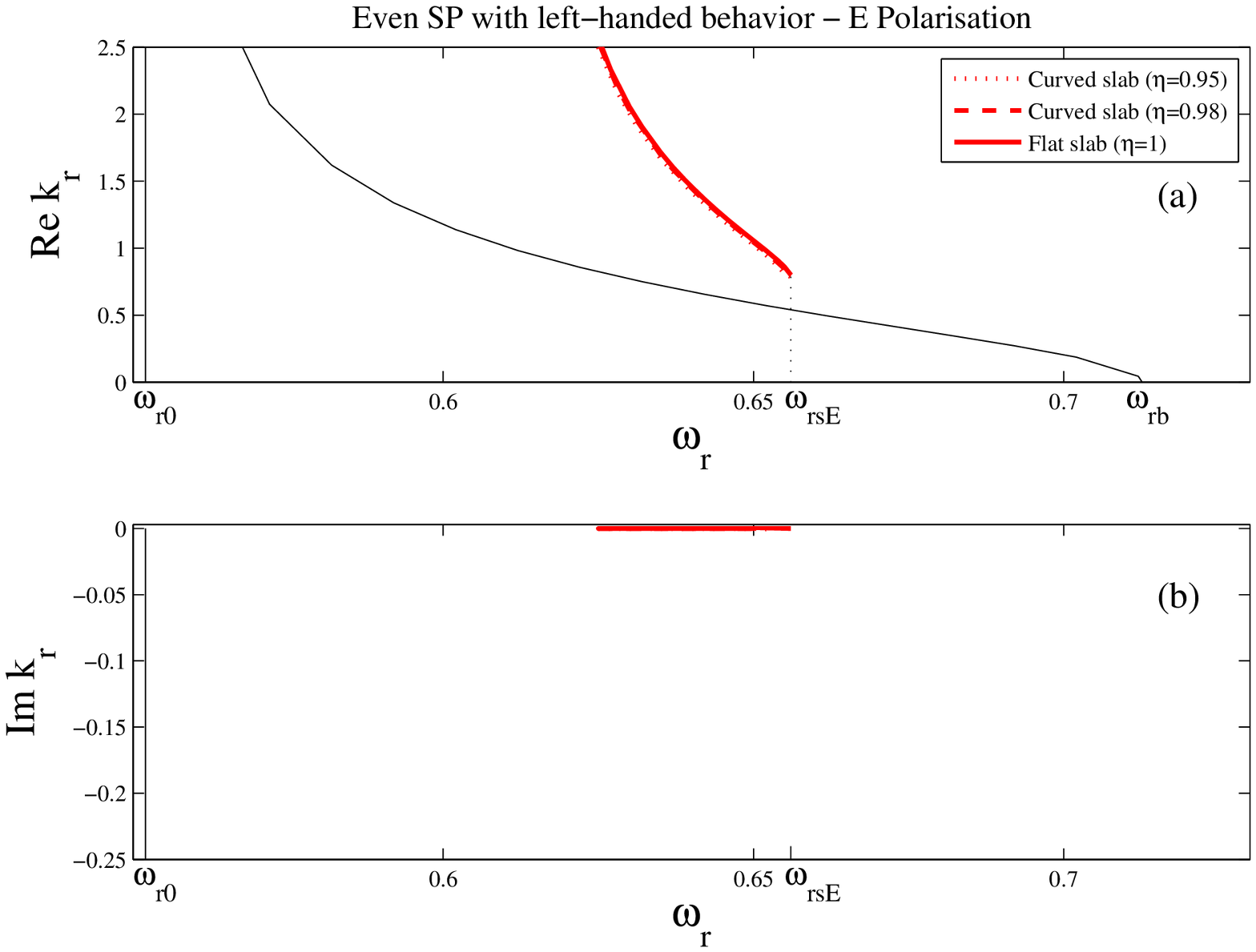}
\caption{Dispersion relations (a) and attenuations (b) of surface
polaritons
guided in left-handed flat and curved slabs embedded in vacuum ($\protect%
\varepsilon _{1}=1,$ $\protect\mu _{1}=1$): role of the slab
curvature. \ Even surface polariton with left-handed behavior, $E$
polarization, frequency range $\protect\omega _{r}>\protect\omega
_{r0}$.} \label{fig:EvenSP_lh_polE_HF}
\end{figure}

\begin{figure}[tbp]
\includegraphics[height=6.5cm,width=8.6cm]{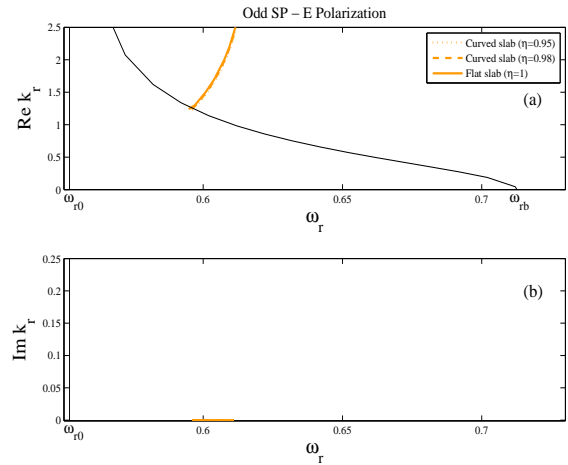}
\caption{Dispersion relations (a) and attenuations (b) of surface
polaritons
guided in left-handed flat and curved slabs embedded in vacuum ($\protect%
\varepsilon _{1}=1,$ $\protect\mu _{1}=1$): role of the slab
curvature. \
Odd surface polariton, $E$ polarization, frequency range $\protect\omega %
_{r}>\protect\omega _{r0}$.} \label{fig:OddSP_polE_HF}
\end{figure}

In Figs. \ref{fig:EvenSP_polH_BF}-\ref{fig:OddSP_polE_HF} we display
the dispersion relations and the attenuations of the surface
polaritons guided in the left-handed cylindrical slab embedded in
vacuum ($\varepsilon _{1}=1$ and $\mu _{1}=1$). Even though we have
restricted ourselves to that configuration, the results we obtained
numerically are in fact very general and they permit us to correctly
illustrate the theory. In particular, the global aspects of the
dispersion and attenuation curves are rather independent of the
value of $\varepsilon _{1}.$ These curves are plotted in the form
$\mathrm{Re}\ k_{r}=\mathrm{Re}\ k_{r}\left( \omega _{r}\right)$ and
$\mathrm{Im} \ k_{r}=\mathrm{Im}\ k_{r}\left( \omega _{r}\right) $
where $k_r$ is the reduced wave number
\begin{equation}
k_{r}=\frac{\lambda d}{ac}  \label{par_red_3}
\end{equation}%
while $\omega _{r}$ is the reduced frequency already defined in Eq. (\ref%
{par_red_2}). The characteristics of the left-handed material are
those previously given in Section 2.D. Here, it is important to note
that (i) we use the same definition of the reduced frequency $\omega
_{r}$ for both the flat and the curved slab and (ii) the case of the
flat slab examined in the previous section can be formally recovered
by taking the limit $\eta \rightarrow 1$ and keeping
$d=(a-b)=\mathrm{const}$. These considerations permit us to compare
on same plots the properties of the surface polaritons guided in
flat ($\eta=1$) and curved ($\eta=0.95$ and $\eta=0.98$) slabs. It
should be also noted that we have kept the terminology
\textquotedblleft even\textquotedblright\ and \textquotedblleft
odd\textquotedblright\ in spite of the symmetry breaking induced by
the curvature of the slab.


In Figs. \ref{fig:EvenSP_polH_BF}-\ref{fig:OddSP_polE_HF} we can see that
the surface polariton dispersion relations change only very little with the
slab curvature. In Figs. \ref{fig:EvenSP_polH_BF}, \ref{fig:OddSP_polH_BF}, %
\ref{fig:EvenSP_rh_polH_HF}, \ref{fig:EvenSP_rh_polE_HF} we can observe that
curvature induces attenuation of the considered surface polaritons. This is
in accordance with the usual behavior of surface polaritons guided by curved
interfaces and it can be interpreted in terms of energy radiated away from
the interfaces. Such a behavior does not occur for the surface polaritons
described in Figs. \ref{fig:EvenSP_lh_polH_HF}, \ref{fig:OddSP_polH_HF}, \ref%
{fig:EvenSP_lh_polE_HF}, \ref{fig:OddSP_polE_HF} where no attenuation is
observed. This is of course very surprising and constitutes the main result
of our study. This is observed in the frequency range $\omega _{r}>\omega
_{r0}$ for the \textquotedblleft odd\textquotedblright\ surface polaritons
and for the \textquotedblleft even\textquotedblright\ surface polaritons
with left-handed behavior.

\section{Conclusion}

In this article, we have described the surface polaritons guided in
a left-handed cylindrical slab. We have shown that the slab
curvature slightly modifies the surface polariton dispersion
relations. It is well known that, in general, curvature induces
attenuation of surface polaritons due to energy radiated away from
the interfaces. However, it is worth pointing out that for the
left-handed cylindrical slab, under certain conditions, surface
polaritons can propagate without loss. This is true, in particular,
for the surface polaritons presenting a left-handed behavior.

Surface polaritons propagating without attenuation are particularly useful
with in mind practical applications in the field of optical communications
as well as the development of photonic integrated circuits and ultra-compact
plasmon-based integrated circuits. It is therefore very interesting to know
that it is possible to transmit information without loss by using
left-handed curved waveguides and that this can be achieved by using surface
polaritons with unusual properties.

Finally, it should be noted that, in this article, we have restricted to
surface polaritons our study of the modes guided in a left-handed curved
slab. It is important to recall that such a waveguide is in fact a very rich
system: indeed, in addition to the surface polaritons considered here, there
also exists an infinite family of oscillating guided modes already present
on the left-handed flat slab as well as an infinity of new guided modes of
whispering-gallery-type which have no analogs in the flat slab case. They
could also play an important role in the context of nanoparticle physics (in
order to fully understand the resonant properties of hollow spheres made of
a negative-refractive-index material) or in the context of optical
transmission of information.

\end{document}